\documentclass{article}

\usepackage{arxiv}

\usepackage[utf8]{inputenc} 
\usepackage[T1]{fontenc}    
\usepackage{hyperref}       
\usepackage{url}            
\usepackage{booktabs}       
\usepackage{amsfonts}       
\usepackage{microtype}      
\usepackage{cleveref}       
\usepackage{graphicx}       
\usepackage[numbers]{natbib} 

\title{Many Processors, Still One Computer:\\The Nested Parallel von Neumann Architecture and Nested BSP}

\date{}

\author{
  Liao Heng \\
  Huawei Technologies Co., Ltd. \\
}

\begin{document}

\maketitle

\begin{abstract}
The contest in large-scale AI computing is no longer whether one processor can be made stronger, but whether a million of them can still behave as a single computer. This paper argues that two extensions are required. First, classic BSP, nested recursively, becomes a plan that holds at every scale: each superstep consists of parallel work, then exchange and aggregation, then a barrier, and at every layer the participating units are peers. Second, the von Neumann single-machine architecture, extended past its master--slave habit, becomes the Nested Parallel von Neumann Architecture: a nest of peer-equal layers, from chip package to autonomous zone, joined end to end by one memory-semantic interconnect, the Unified Bus. The software nest and the hardware nest correspond layer by layer, and the $\tau$ Scaling law folds time at every layer. We examine which workloads the nested structure serves, AI training above all but much of classic HPC as well, and describe the hardware decisions that make the nesting physical. Many processors, still one computer.
\end{abstract}

\section{Introduction: The Question Eighty Years Left Open}

One question now sits beneath every decision in large-system design: when processors number a hundred thousand, or a million, is it still clear how to design one?

How to build a computer, von Neumann made clear \cite{vonneumann1945}. For eighty years the industry followed that path, making processors stronger generation after generation. How to turn so many processors into one coordinated whole he never said. That is no fault of his: he was describing one machine, not a campaign of a million.

Chip contests used to look like two martial artists sparring, where the harder punch wins. An AI campaign is not fought that way. A million soldiers take the field, and what matters is not who hits hardest but whether the formation holds. At a hundred thousand or a million, victory turns on a single question: how to bind the processors into one system and, once bound, whether the result is still one computer.

A formation holds together only with two things: a plan of action and a chain of command that reaches every unit. Without a clear plan, and without orders that get through, numbers alone are sand. When a million units must move at once, those orders are hard to give.

Two extensions follow, and they are the subject of this paper. The first belongs to software: classic BSP, nested recursively, under one rule. At every layer, the participating units are peers. The second belongs to hardware: the von Neumann single-machine architecture extended into the Nested Parallel von Neumann Architecture, from chip package to autonomous zone, joined end to end by one memory-semantic interconnect, the Unified Bus. The two extensions are nesting dolls, and they must fit.

\section{From Turing to von Neumann: A Programmable Procedure and a Programmable Machine}

Two foundational contributions are often compressed into one story; before asking how to program a million processors, we separate them here.

Turing's theory is a theory of computation and programmability \cite{turing1936}. His 1936 machine is minimal: a finite control with a finite number of states, an unbounded tape of symbol cells, and a finite table of rules. Each rule reads the current state and the symbol under the head, then writes a symbol, moves the head one cell, and enters a next state. That finite table is the program, and every step it prescribes is drawn from one fixed, finite repertoire of elementary operations. A computation is therefore a finite description composed from a fixed set of primitive operations, applied step by step to storage.

The phrase ``Turing-complete instruction set'' requires care, because the term names a property of a system rather than a badge on a particular list of opcodes. An instruction set is Turing-complete when any computable function can be expressed as a finite program over that set; equivalently, the set is expressive enough to simulate a universal Turing machine. Three capabilities are what matter:

\begin{enumerate}
\item \textbf{Data-dependent control flow.} The program must be able to choose its next step according to a value it has computed: a conditional branch, or any construct from which one can be built. Without it, a program is a fixed-length script rather than an algorithm.
\item \textbf{Unbounded addressable storage.} The machine must be able to use as much memory as a computation requires. Turing's tape has no end; no physical machine has this property, so a real computer is, strictly speaking, a very large finite-state machine rather than a Turing machine. Turing completeness is always asserted on the assumption of unbounded storage.
\item \textbf{Composition and repetition.} Elementary operations must be combinable without limit, so that arbitrarily long computations can be described by a finite program.
\end{enumerate}

This definition has two consequences. First, Turing completeness says nothing about how many instructions a set contains, nor how convenient or fast they are: single-instruction sets are Turing-complete, and richer instruction sets buy efficiency and expressiveness rather than additional computational power. Second, and more consequential for architecture, the universal Turing machine shows that a machine's program can itself be written on the tape as data and interpreted by another machine. Programs and data are the same kind of object. One general-purpose machine can therefore run any algorithm without being rebuilt for each one, which underlies everything now called software.

What Turing did not supply was a machine. A Turing machine says nothing about how to arrange memory, arithmetic, control, and input/output so that the steps run quickly, or run at all, in electronics. The von Neumann architecture supplied exactly that organization \cite{vonneumann1945}, and it did so by making the universality result physical. Its stored-program principle places instructions and data together in one addressable memory; a control unit fetches, decodes, and executes those instructions in sequence, tracked by a program counter; an arithmetic and logic unit performs the operations; input and output connect the machine to the world. The fixed repertoire of primitive operations becomes a concrete instruction set architecture with opcodes encoded in bits, and the endless tape becomes a finite, addressable memory hierarchy. The program is no longer wired into a special-purpose machine: change the contents of memory and the same hardware performs a different computation.

\begin{figure}[htbp]
\centering
\includegraphics[width=\linewidth]{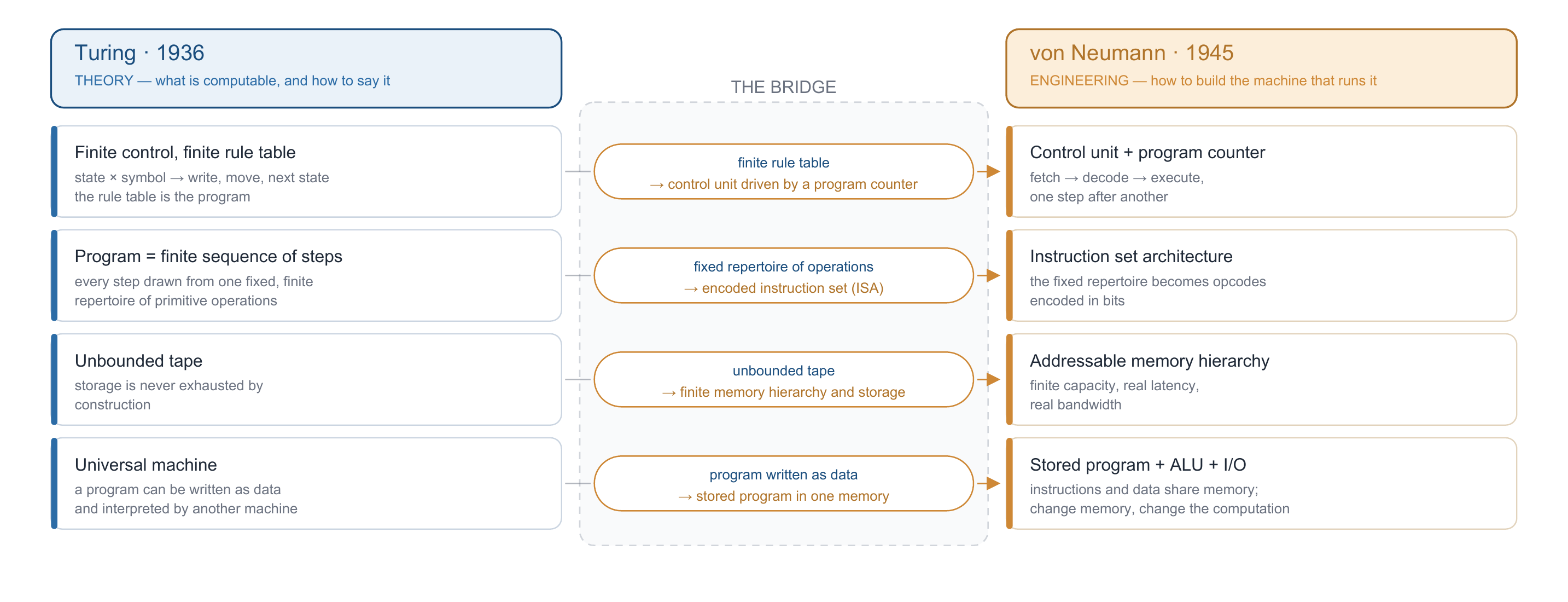}
\caption{From Turing to von Neumann: crossing from a theoretical concept to an engineering practice. Turing settled how a computation is specified to a machine (finite control and rule table, a program as a finite sequence of primitive steps, an unbounded tape, a universal machine); von Neumann settled how to build the machine that performs it (control unit with program counter, instruction set architecture, addressable memory hierarchy, stored program with ALU and I/O). Each theoretical construct maps onto an engineering counterpart, shown as the bridge in the middle.}
\label{fig:turingvn}
\end{figure}

The relationship is complementary rather than competitive:

\begin{itemize}
\item \textbf{Turing:} What is a computable procedure, and how can that procedure be encoded as a program?
\item \textbf{Von Neumann:} How can a practical stored-program machine be built to execute such programs?
\item \textbf{BSP:} How should a program be structured when the machine contains many processors working concurrently?
\end{itemize}

The Turing abstraction does not require von Neumann hardware, and not every modern computer is a textbook von Neumann machine. Yet the pairing proved effective. The universal-program idea supplied the logic, and the stored-program architecture supplied the physical machine.

The difficulty in the AI era is that this pairing was developed around a predominantly sequential view of execution. A von Neumann computer can contain parallel units, but its classical form does not by itself explain how a million processors should coordinate as one programmable system. That is the gap BSP begins to address, and the gap that Nested BSP and the Nested Parallel von Neumann Architecture set out to close across multiple levels of hierarchy.

\section{The Classic Answer: BSP and the Flat Supercomputer}

The high-performance computing (HPC) community encountered this coordination problem long ago, and its answer is where any discussion of million-processor programming has to begin.

\subsection{BSP: How to Program One Large Parallel Computer}

Leslie Valiant introduced the Bulk Synchronous Parallel (BSP) model \cite{valiant1990}, with Bill McColl contributing substantially to its development and use \cite{mccoll1995}. Where Turing and von Neumann had settled how to specify and build a machine that executes one instruction stream, BSP asked the next question: what is the right way to write a program when the machine contains many processors that run at the same time?

BSP answers with a simple abstraction of the machine (processors that each hold local memory, an interconnect that delivers messages between them, and a mechanism that synchronizes all of them), together with an equally simple structure for the program. A BSP computation proceeds as a sequence of \emph{supersteps}, and each superstep has three parts:

\begin{enumerate}
\item every processor computes on data held in its local memory;
\item processors exchange the data their peers will need;
\item a barrier ends the superstep, guaranteeing that its communication is complete and visible before the next superstep begins.
\end{enumerate}

The discipline is what makes the model valuable. The programmer no longer reasons about arbitrary interleavings of instructions and arrivals of messages; correctness is argued phase by phase. BSP is also a cost model. A machine is characterized by a small number of parameters (the number of processors, the rate at which the interconnect can deliver traffic, and the latency of a barrier), and the cost of a superstep is the slowest local computation, plus the communication volume divided by network throughput, plus the synchronization cost. That is what lets an algorithm designer predict how a program will behave on a given machine.

\subsection{The Machine BSP Described: The Flat Supercomputer}

The machine BSP had in view was the massively parallel processor (MPP), and the commodity and accelerated clusters that inherited its structure: many nodes, each with its own processors and local memory, joined by a dedicated high-performance interconnect, with no single shared memory spanning the machine. The programming abstraction placed on top of that hardware was essentially \emph{flat}: one global set of peer ranks, one address space per rank, one network assumed to behave roughly the same between any two of them, and one global barrier that the whole machine enters together.

Flatness was a reasonable bargain. At thousands of nodes, running one dense simulation, with an interconnect engineered to be as uniform as possible, describing the machine with a single throughput parameter and a single barrier latency was close enough to the truth to be useful, and it gave the field a portable way to reason about parallel programs.

But the physical machine was never actually flat. Two cores in a package, two chips on a board, two racks in a row, and two racks across a data hall differ by orders of magnitude in bandwidth and latency. The flat abstraction did not remove that hierarchy; it hid it, and the cost was paid in performance.

\subsection{How Supercomputers Are Programmed in Practice}

BSP and the tools practitioners actually use operate at different levels, and the relationship between them must be stated precisely. BSP is a programming and performance model: a way of structuring computation and predicting its cost. MPI, OpenMP, and accelerator programming systems are mechanisms through which such a structure is implemented.

\begin{itemize}
\item \textbf{MPI} is a distributed-memory programming interface. Point-to-point operations carry communication; MPI\_Barrier provides synchronization; and collectives such as broadcast, gather, and Allreduce implement the exchange-and-aggregate patterns that BSP supersteps require. An MPI program can follow BSP supersteps faithfully, but MPI also permits asynchronous and irregular programs that are not BSP at all.
\item \textbf{OpenMP} is a shared-memory programming model. Parallel regions, work-sharing loops, tasks, reductions, and barriers express concurrency among the cores and threads inside one node.
\item \textbf{Accelerator models and collective libraries}, such as CUDA, SYCL, and the device collective libraries that accompany them, express massively parallel work within a device and synchronization or reduction across devices.
\item \textbf{PGAS and shared-address models} such as SHMEM, UPC, and Fortran coarrays take a third route: one global address space with explicit awareness of locality, closer in spirit to memory semantics than to message passing.
\end{itemize}

In practice, no large machine is programmed with only one of these. The prevailing style is hybrid, often written \emph{MPI + X}: MPI coordinates nodes, OpenMP coordinates cores within a node, accelerator kernels coordinate thousands of lanes on a device, and a collective library connects the accelerators. Each mechanism is strong at its own level. But they are assembled as largely separate programming worlds, each with its own assumptions about memory, synchronization, and failure, and the boundaries between them are exactly where performance and reasoning break down.

\subsection{Where the Flat Picture Strains at AI Scale}

At a million processors, every place where flatness diverged from physics becomes a first-order problem.

A single global barrier must now be entered by participants separated by orders of magnitude in distance, so its cost is set by the slowest and farthest, and a collective over the whole machine crosses every boundary in the hierarchy. One throughput parameter and one latency parameter can no longer describe a machine that spans a package and a data center. Component failure stops being an exception: in clusters of tens of thousands of accelerators, a lost link or module is routine, and a single flat synchronization domain propagates that loss to everyone. Underneath all of it sits an assumption inherited from single-machine design, a host that commands and devices that obey, and this assumption turns the center into a bottleneck precisely when the system grows.

None of this makes BSP wrong; it makes BSP \emph{incomplete} for this era. The structure of local work, communication, and synchronization is exactly right; what is missing is hierarchy, and peer equality within each level of that hierarchy.

\section{Nested BSP and the Nested Parallel von Neumann Architecture}

\subsection{Nested BSP: An AI-Era Extension}

The extension proposed here is \textbf{Nested BSP}: classic BSP, nested recursively.

The idea is simple. The outermost layer is BSP for the whole system. Each unit of local work at that layer may itself be a parallel BSP computation at a smaller scope: local work, communication, barrier, then the next superstep. Inside that is another layer, and then another, in recursion all the way down. A global training step may contain pipeline stages; a stage may contain expert or data-parallel groups; a group may contain tensor-parallel operations; and an accelerator may execute thousands of parallel lanes. Each level has its own locality, communication cost, and synchronization boundary.

Where classic BSP describes one flat machine with one set of cost parameters, Nested BSP describes a hierarchy of machines: every layer is a BSP machine in its own right, with its own throughput, its own barrier latency, and its own scope of synchronization. Instead of hiding the hierarchy, the model names it.

What Nested BSP adds beyond that hierarchy is one architectural constraint: at every layer, the participating units are peers, with no master through which every barrier, transfer, or reduction must pass and no slave that can only obey; parallelism is nested, but authority is not.

\begin{figure}[htbp]
\centering
\includegraphics[width=0.55\linewidth]{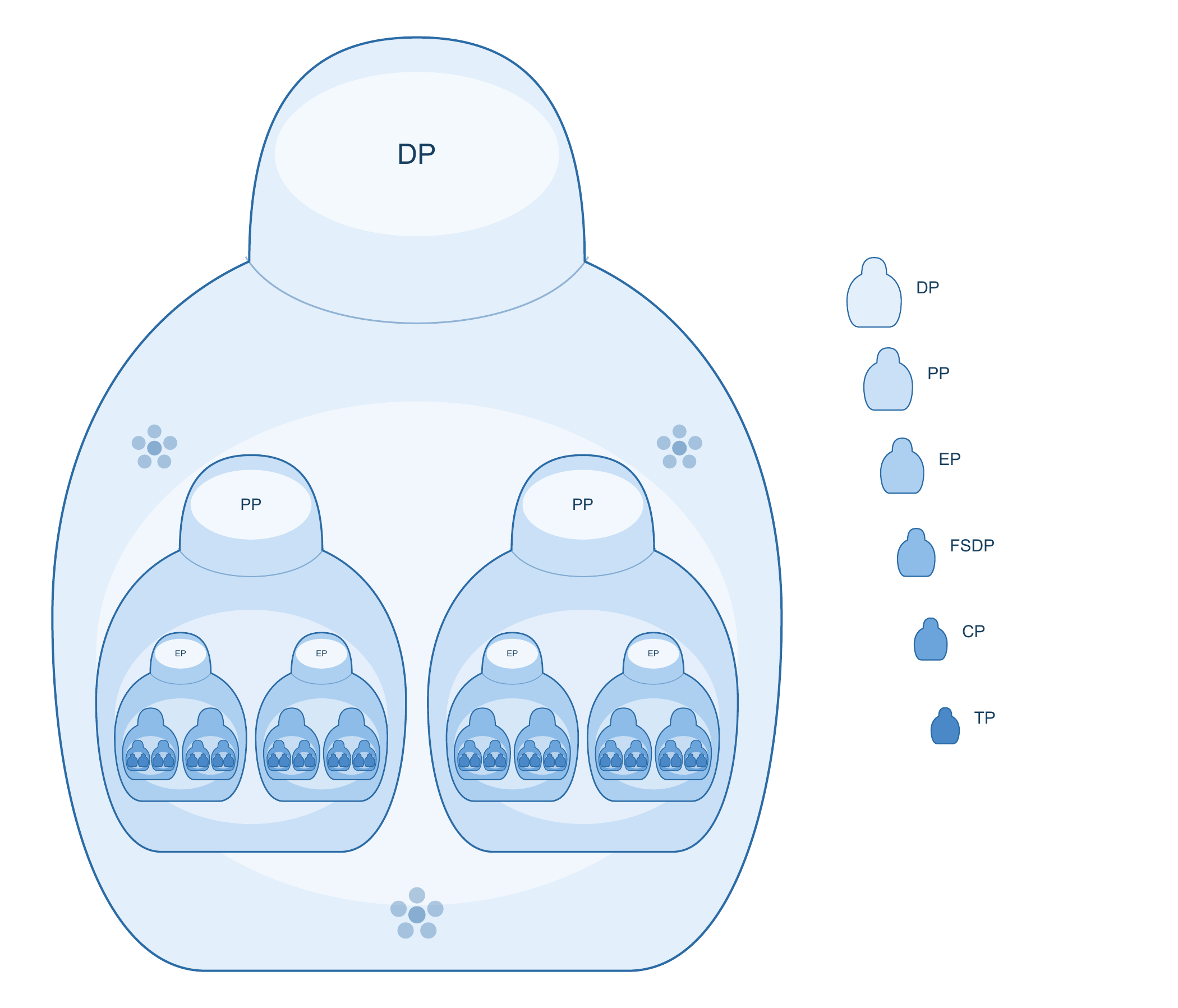}
\caption{Software nesting dolls: Nested BSP. From outermost to innermost, the levels are data parallel (DP), pipeline parallel (PP), expert parallel (EP), fully sharded data parallel (FSDP), context parallel (CP), and tensor parallel (TP).}
\label{fig:sw}
\end{figure}

The software nest looks like this (Fig.~\ref{fig:sw}): six parallelism dimensions, from the machine-wide DP at the outermost layer to the intra-operation TP at the innermost, each doll a BSP layer with its own scope.

On today's AI systems, the layers may include data parallelism (DP), pipeline parallelism (PP), expert parallelism (EP), fully sharded data parallelism (FSDP), context parallelism (CP), and tensor parallelism (TP). Their exact nesting depends on the model and the implementation; no universal order should be assumed. But each can be understood through the same disciplined pattern: perform work concurrently within a defined group, exchange or reduce the required state, synchronize where correctness requires it, and then advance. A problem at machine scale is decomposed into problems that each layer can manage.

One detail is easy to miss: inside each larger scope there is not one smaller unit, but \emph{many}. Depending on the parallelization plan, an outer group is partitioned into multiple pipeline stages, expert groups, shards, tensor-parallel ranks, or other inner groups. These dimensions may be nested or composed rather than arranged in one fixed order. That multiplicity is the resource parallelism draws on.

A plan of this shape needs a machine of the same shape, and that machine is the second extension.

\subsection{A Machine of the Same Shape: The Nested Parallel von Neumann Architecture}

Huawei's SuperNode cluster design rests on Nested BSP as its theoretical foundation. Everything that follows about hardware (one protocol end to end, memory semantics, full peer equality, copper near and optics far) exists so that this nested structure can stand in the physical world. That hardware nest is the Nested Parallel von Neumann Architecture: from package to autonomous zone, concurrent units at every layer, peers on one bus.

The hardware is also a nest of layers, each with its own boundary. Counting from the inside out: core, chiplet, chip package, board, rack, SuperNode, data hall, autonomous zone, data center, and, outermost, inter-data-center.

\begin{figure}[htbp]
\centering
\includegraphics[width=0.55\linewidth]{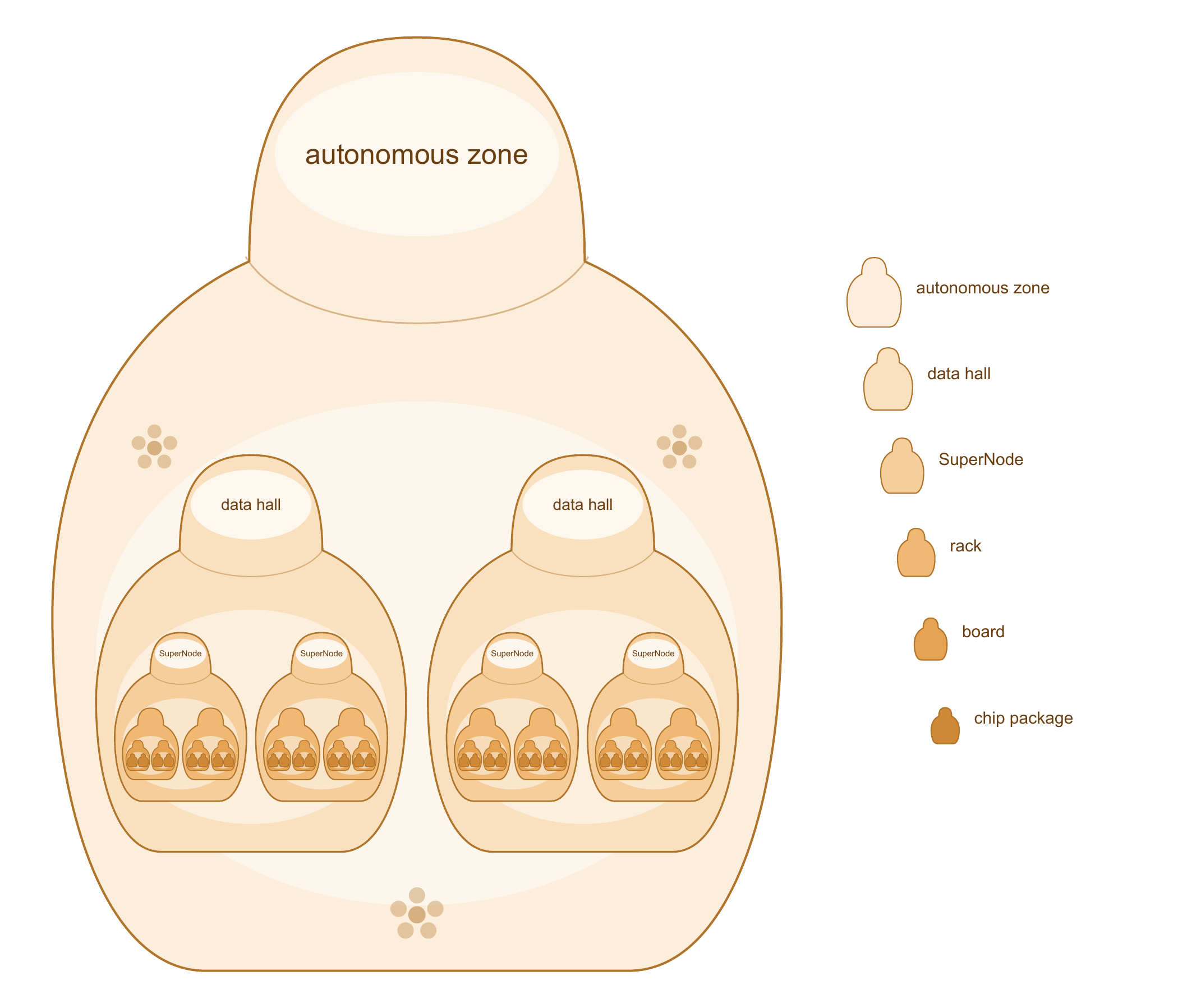}
\caption{Hardware nesting dolls: the body of the machine, from chip package to autonomous zone. One autonomous zone holds many data halls, one hall holds many SuperNodes, and one SuperNode holds many racks, all the way down to the package.}
\label{fig:hw}
\end{figure}

The hardware layers nest one within another, each holding many of the next, from the package to the autonomous zone (Fig.~\ref{fig:hw}). The software nest above and this hardware nest must line up, layer by layer.

The interesting stretch is not the innermost transistor or the outermost campus, but the middle of hardware system design: from the chip package to the cluster that fills a data center. Further in is how to train one soldier; further out is how campuses talk to campuses. How a hundred thousand or a million become one system is decided in these middle layers (package, board, rack, SuperNode, data hall, autonomous zone), where command must pass down intact. At each boundary the question repeats: whether it is still one plan, one command set, one computer.

The Unified Bus exists to answer it. It makes the split and join of every layer line up, so that command does not break at the boundaries.

\subsection{$\tau$ Folds Time at Every Layer}

At the system level, the $\tau$ Scaling law \cite{he2026tau} maps onto these two nests.

At every Nested BSP layer and every physical scale, $\tau$ does the same thing: it folds time. The playbook at every layer is identical. Work that would have been done one piece after another is spread across many units that can work at once on that layer, and that layer's time constant folds shorter. The folding repeats once inside the package, once on the board, once in the rack, once in the SuperNode, and once again in the data hall.

The software nest is Nested BSP; the hardware nest is the Nested Parallel von Neumann Architecture, with more concurrent, peer-equal units at every layer, in the package, on the board, in the rack, and in the hall. The two nests correspond layer by layer, and time folds at each. One layer alone folds only so much. Six layers fold multiplicatively, and the time of the whole computation comes down.

\subsection{The Two Nests Must Fit}

The two nests must fit with no gap. Where a layer fails to match, the step time scatters, whether in the protocol, in the software, or in the network between racks. No matter how many parallel units are placed, they cannot yield the intended performance.

The nest follows the same recursive logic as the physical world: farther out, larger scale, longer distance, lower bandwidth, higher latency; farther in, smaller scale, shorter distance, higher bandwidth, lower latency.

This is a continuation of von Neumann's idea, not a break from it: a hierarchically nested parallel computing system in which the principle does not change. What changes is how large the computer is, how many layers it nests, and how far a single command can reach: distant memory must feel as natural as memory at hand. Chips are soldiers. The computer is the whole formation, from package to autonomous zone, answering to one command set.

\section{Programming the Nest: What Fits, and What It Takes}

With the two nests aligned, what decides whether any of it matters is what a program looks like on such a machine: which programs can take this shape at all, and what it costs to write them.

\subsection{AI Training Already Writes Nested BSP by Hand}

Large-model training is the one workload that arrived at this structure on its own. Every parallelism dimension in use today is a BSP layer with its own scope, its own communication pattern, and its own synchronization frequency:

\begin{itemize}
\item \textbf{Tensor parallelism (TP)} splits individual matrix operations across a small group. It synchronizes many times per layer, per token step, with reductions on the critical path. It is the innermost, most frequent, and most latency-sensitive layer.
\item \textbf{Context or sequence parallelism (CP)} splits the sequence, exchanging key/value blocks in ring or neighbor patterns. The traffic is bandwidth-hungry, but its communication structure stays local.
\item \textbf{Expert parallelism (EP)} dispatches tokens to experts and combines the results. Its all-to-all traffic is bursty and sensitive to bisection bandwidth rather than to latency alone.
\item \textbf{Fully sharded data parallelism (FSDP)} all-gathers parameters before use and reduce-scatters gradients after. The volume is high but schedulable, and it overlaps with compute.
\item \textbf{Pipeline parallelism (PP)} hands activations and gradients between stages point to point. It is the most latency-tolerant layer, which is why it survives across racks.
\item \textbf{Data parallelism (DP)} reduces gradients once per optimizer step. It is the least frequent synchronization of all, and therefore the one that can span halls.
\end{itemize}

From the bottom up, the pattern is consistent: synchronization becomes rarer and coarser as scope widens. That is precisely the property Nested BSP formalizes and that the nested hardware is built to exploit: frequent, tight supersteps confined inside one hop, infrequent ones allowed to travel. Practitioners express this today with device meshes and process groups: sub-communicators carved by hand out of a flat rank space, with the mapping onto packages, boards, racks, and halls maintained informally in a configuration file. The structure is already nested; what is missing is a model and a machine that treat that nesting as a first-class object rather than as a tuning artifact.

\subsection{The Test: Can the Problem Be Cut Into Smaller Copies of Itself?}

The AI case suggests a single criterion, and it decides which workloads belong in this discussion at all.

AI workloads fit the nested model because they tile recursively: a tile of a matrix multiply is again a matrix multiply, a block of attention is again attention, and a shard of a layer is again a layer. At every level of the nest (lane, package, board, SuperNode) the program is solving the same kind of problem at a smaller scale, and the results are joined by an aggregation whose footprint is small compared with the work that produced it. That is why an AI program can be written once and mapped at many scales.

The test for any workload is therefore this: at any scope, can the problem be partitioned into a few problems of the same kind at smaller scale, joined by an exchange or aggregation whose cost is small relative to the local work? If yes, the workload fits Nested BSP, and the nested machine can serve it at every layer. If no, that is, if the only honest description is one flat problem over all participants, then the workload does not fit the nested model, and no amount of hierarchy in the hardware will make it scale.

Two things make the difference. The first is \emph{surface to volume}. Double the edge of a three-dimensional subdomain and the local work grows eightfold while the halo grows fourfold; double the edge of a matrix tile and the arithmetic grows eightfold while the data moved grows fourfold. Recursion pays precisely because the interface between subproblems grows more slowly than the work inside them, so each layer inward is more self-sufficient than the layer outside it. The second is a \emph{bounded interface}: a halo, an $R$ factor, a frontal matrix, a multipole expansion, a gradient sum. When what crosses the boundary can be summarized in a small object, the recursion closes. When every unit needs a little of everything from every other unit, it does not.

Most physical interactions weaken with distance: gravitational and electrostatic fields fall off as $1/L^2$, radiated intensity likewise, diffusive and screened interactions faster still. The influence on any region is dominated by its near neighbors, while the far field can be summarized rather than resolved in detail. That summary is exactly a bounded interface, and it is why multipole expansions, coarse grids, and halo exchanges work at all. Many physical systems are also self-similar across scales, the same structures recurring from the small to the large. A computational model whose layers are copies of one another is therefore not an artificial construction imposed on science; it matches the geometry of the problems themselves. That is why so much of high-performance computing, including fields, particles, meshes, and the solvers built on them, is natively hierarchical, and why a nested program on a nested machine is a natural fit rather than a forced one.

One caveat matters, because this is where most of the interesting engineering lies: recursive decomposability is a property of the \emph{algorithm}, not of the problem. Direct summation of $N$-body interactions is flat; the fast multipole method solves the same physics recursively. A classical two-sided SVD is latency-bound and hard to nest; randomized SVD is a matrix multiply, a tree reduction, and a small dense solve. Dense LU with partial pivoting has a latency chain along a process column; communication-avoiding pivoting replaces it with a tournament, that is, a tree. In nearly every case where a workload is called a poor fit, the accurate statement is that no recursive formulation is known \emph{yet}.

\subsection{Classic HPC Workloads Under Nested BSP}

Applying that test to the classic HPC codes is instructive. These are the workloads BSP was invented for, and they were written for the flat machine. Suppose their authors were willing to reprogram them against a nested model: little of the \emph{mathematics} would have to change, because the algorithms were already recursive; it was the flat programming model that forced them to pretend otherwise.

\paragraph{Dense direct factorization --- HPL, and LU, Cholesky, QR generally.}
HPL is a blocked LU factorization over a two-dimensional block-cyclic distribution. Each step factors a panel, searches for pivots, broadcasts the panel, and updates the trailing submatrix. Those pieces call for radically different layers. The trailing update is a large matrix multiply with high arithmetic intensity. It belongs inward, where bandwidth is cheapest, and it should be nested again into board-level, package-level, and lane-level tiles. Pivot search is a reduction along a process column with a small payload and a long dependency chain. It is pure latency, and it must be confined within one-hop reach; if it crosses a rack boundary it sets the pace of the whole factorization. The panel broadcast is comparatively infrequent and can be allowed to travel outward. In a flat model these three live in one rank space with one cost parameter; under Nested BSP each is placed at the layer whose cost it can afford. The field's hardest-won optimizations, including recursive blocking, tile algorithms, communication-avoiding pivoting, 2.5D matrix multiplication, and TSQR, are all reformulations that trade synchronization frequency for locality. They are hierarchy-aware algorithms written in a model that cannot say the word ``hierarchy.'' In a nested model they become the natural expression rather than exceptional measures.

\paragraph{Iterative solvers --- HPCG, Krylov methods, multigrid, domain decomposition.}
HPCG is the counterpart to HPL that avoids inflated peak rates: sparse matrix--vector products with halo exchange, global dot products, and a multigrid preconditioner with a Gauss--Seidel smoother. Nothing here is compute-bound; everything is bound by memory bandwidth and synchronization latency. The mapping under Nested BSP is straightforward. Halo exchange is nearest-neighbor traffic and belongs at an inner layer, where the neighbors are physically near. Global dot products are reductions over the entire machine and belong at the outermost layer; because Krylov methods need one per iteration, reducing their count is exactly what $s$-step and pipelined variants do: fewer outer supersteps, more local work each. The most telling case is the multigrid V-cycle, which is itself a hierarchy: fine levels hold abundant parallelism and local communication; coarse levels hold almost no work and consist of little but synchronization. On a flat machine the coarse-grid solve is the classic scalability wall, because a tiny problem is spread across the whole machine and pays global latency for every step. Under a nested model the right move is available and expressible: agglomerate the coarse levels inward --- fold them into a single one-hop domain, a SuperNode or even a package, where a barrier costs nanoseconds instead of microseconds --- while the fine levels stay spread out. Multilevel domain decomposition (Schwarz, FETI, BDDC) has the same shape: subdomains within subdomains, with coarse spaces that belong at progressively inner layers.

\paragraph{Sparse direct solvers.}
Here the fit is literal. Nested dissection ordering produces an elimination tree; the multifrontal method processes independent subtrees in parallel and merges their frontal matrices at the parent; the community has long mapped subtrees to sub-machines, even calling the dense blocks in the tree ``supernodes.'' That is Nested BSP with the names changed: each subtree is a BSP computation at an inner scope, each merge is the parent layer's exchange-and-aggregate, and recursion terminates in the leaves. Near the root, a few very large dense fronts need the aggregated bandwidth and memory of an outer layer; near the leaves, thousands of small fronts need locality and nothing else. The nested machine can serve both; a flat machine must choose.

\paragraph{SVD, PCA, and eigenproblems.}
These decompose into one recurring pattern: reduce inward, solve a small problem in one hop, broadcast outward. Tall-skinny QR is a tree reduction of $R$ factors, reduced within the package, then the board, then the rack, then the hall. That is a textbook nested reduction, and it is why randomized SVD scales so well: it is a large matrix multiply, a tree reduction, and a tiny dense decomposition. PCA has the same skeleton: forming a covariance or Gram matrix by a reduction over samples, then solving a comparatively small eigenproblem, which under a nested model should be agglomerated inward rather than distributed. Full-accuracy SVD and symmetric eigensolvers are harder, because two-sided reductions carry long latency-bound panel chains; the field's answer, two-stage band reduction (which converts most of that work into matrix multiply) together with divide-and-conquer for the banded or tridiagonal problem, is again recursion and again layer placement. Divide-and-conquer \emph{is} nesting: the split is an outer superstep, the subproblems are inner BSP computations, the merge is the aggregation.

Across all of these, what the nested hardware contributes is what makes the inner layers worth having at all. Memory semantics bring a communication round trip to roughly a hundred nanoseconds, so an inner superstep is cheap enough to be worth creating. One-hop reach (the product of switch radix and chip radix, discussed in Section~\ref{sec:practice}) decides how much of an algorithm's frequent synchronization fits inside a single hop, which is to say how far inward the pivot search, the coarse-grid solve, or the small eigenproblem can be folded. Peer equality means a reduction can be initiated without funneling through a host. And $\tau$ folds time at each layer, so the gains multiply rather than add. The cost is real: these codes must be reprogrammed to say which scope each phase belongs to. The return is that the hierarchy-aware algorithm variants people already know stop fighting the programming model.

\subsection{What Fits, and What Does Not}

Applying the test across the workloads that matter gives a rough but useful assessment (Table~\ref{tab:fit}).

\begin{table}[htbp]
\centering
\caption{Classic HPC and AI workloads under the nested test. For each workload: whether it admits a recursive decomposition into smaller copies of itself, where it strains under nesting, and the overall fit to Nested BSP. Workloads marked weak should be confined to one SuperNode rather than spread across the machine.}
\label{tab:fit}
\small
\begin{tabular}{p{0.21\linewidth}p{0.26\linewidth}p{0.28\linewidth}p{0.14\linewidth}}
\toprule
Workload & Recursive decomposition & Where it strains & Fit \\
\midrule
Dense GEMM, tile LU / Cholesky / QR & A subblock is the same problem; tiles recurse to lane level & Pivot search is a latency chain unless tournament pivoting is used & Strong \\
HPL & Blocked LU: panel, broadcast, trailing update, repeat & Panel and pivoting must stay inside one hop & Strong \\
Structured-grid PDE, domain decomposition, AMR & Subdomain within subdomain; halo is the interface & Deep halos and irregular boundaries at coarse granularity & Strong \\
Multigrid, HPCG & The V-cycle is itself a hierarchy of scales & Coarse levels lose parallelism; must be agglomerated inward & Strong at fine levels, inward at coarse \\
Krylov solvers (CG, GMRES) & SpMV and halo recurse; reductions are associative trees & One global reduction per iteration; $s$-step variants reduce the count & Strong \\
Sparse direct, multifrontal & Nested dissection yields an elimination tree & Tree is unbalanced; root fronts need outer-layer bandwidth & Strong, needs dynamic mapping \\
TSQR, randomized SVD, PCA & Tree reduction plus a small dense core & Full-accuracy two-sided reduction keeps a latency chain & Strong \\
FMM, Barnes--Hut $N$-body & Octree with multipole expansions as bounded interfaces & Load imbalance in adaptive trees & Very strong \\
Molecular dynamics, short range & Spatial decomposition recurses & Long-range electrostatics falls back on FFT & Strong \\
Monte Carlo, ensembles, parameter sweeps & Partition the samples; one reduction at the end & Nothing structural & Very strong \\
\midrule
3D FFT & Cooley--Tukey recursion, pencil or slab decomposition & The transpose is an all-to-all whose volume does not shrink with nesting & Partial: keep the all-to-all in one hop \\
Large-scale sort, shuffle, hash join & Sample or radix partitioning recurses & Full-volume all-to-all at every level & Partial \\
Branch-and-bound, discrete-event simulation & The search or event tree recurses & Global bounds and event order need frequent shared state & Partial; needs peer equality and work stealing \\
\midrule
Graph traversal on scale-free graphs & No good separators; surface is as large as volume & Fine-grained irregular access; recursion saves little traffic & Weak: confine to one SuperNode \\
Sequential recurrences, time stepping, SpTRSV, Gauss--Seidel sweeps & Only the work inside one step decomposes & The dependent dimension does not partition; parallel-in-time is an approximation & Weak beyond the inner layers \\
\midrule
AI training and inference & Tiles, shards, stages, and groups all recurse & Collective volume at the outer layers & Very strong \\
\bottomrule
\end{tabular}
\end{table}

The rows marked weak have a constructive answer rather than a dismissal. A workload that is genuinely flat should be given the largest single tightly coupled domain available, namely one SuperNode: one protocol, memory semantics, one-hop reach, and a barrier in microseconds. Inside that domain, flatness costs comparatively little; across it, flatness costs everything. The practical rule is therefore simple. If the problem nests, spread it across the layers and let each layer pay only the cost it can afford. If it does not nest, fit it inside one SuperNode and do not try to spread it further.

This is also where the scaling arithmetic turns in favor of workloads that were never the target. SuperNodes are growing because AI training and inference demand it, and that demand comes from an industry far larger than high-performance computing has ever been. Every increase in SuperNode size (more nodes in one hop, more aggregate memory, a barrier that stays under ten microseconds at greater scale) raises the ceiling for flat problems that can only live inside one such domain. HPC does not have to fund that growth to benefit from it.

\subsection{Implementing Nested BSP with Today's Mechanisms}

Nested BSP is not a replacement for the mechanisms described earlier; it is the structure that gives the hybrid stack one recursive logic instead of several disconnected ones. An outer superstep can be implemented with MPI across data halls or nodes; its local work can be an OpenMP computation across cores; that work can invoke accelerator kernels and device collectives at still smaller scales. The objective is not to force every application onto a new API, but to align the mechanisms already in use with explicit nested scopes, so that locality, synchronization, communication, and failure containment are each handled at the layer where they belong.

The practical difference is direct. Treating one million processors as a flat MPI communicator makes global barriers and collectives progressively more expensive. Treating a cluster as one shared-memory machine ignores physical distance and coherence cost. Nested BSP preserves a single programming logic while respecting hierarchy: synchronize locally where possible, aggregate within a layer, and communicate outward only when the next layer genuinely requires it.

\subsection{Why MPI + OpenMP Is Less Than Ideal}

That compromise carries a program only so far. MPI + OpenMP offers two abstractions at two levels, across nodes and within a node, for a machine whose meaningful boundaries run core, chiplet, package, board, rack, SuperNode, data hall, autonomous zone. And the two have nothing in common. MPI assumes private address spaces, explicit messages, and a flat space of ranks; OpenMP assumes one shared address space, threads, and implicit sharing. Neither is a restriction of the other, so the boundary between them is not a nesting but a change of language. Accelerators then arrive as a third model bolted onto the second, with the master--slave assumption built in.

The practical costs follow. The decomposition is expressed twice in two vocabularies and kept consistent by hand. Synchronization is also expressed twice, with no single cost model spanning it. And the mapping is frozen into the source, so moving a coarse-grid solve one layer inward is a rewrite rather than a parameter change.

The deeper limitation is the machine these tools describe. A flat supercomputer is a machine with one seam, and flatness does not scale easily to millions of processors: a single global barrier and full-machine collectives grow more expensive with every participant and every boundary crossed, and one throughput parameter with one barrier latency cannot describe a machine spanning a package and a data center. The seam has also moved. When eight thousand nodes share one protocol, one address-space discipline, and a full-machine barrier under ten microseconds, ``inside the node'' and ``outside the node'' is no longer the distinction around which a program should be organized. The mismatch is structural: the machine now has many seams, and they are nested.

\subsection{Why PyPTO Fits: A Nested Machine Model in the Runtime}

The AI software ecosystem met a smaller version of this wall first. Calling a model as a long chain of individual operators became untenable once latency mattered: every call pays launch overhead and moves data through memory it should never have touched. The response was fusion, of which FlashAttention \cite{flashattention2022} and the kernels that followed are examples, and then the recognition that the fused object had grown large enough to need a language of its own. Hence the current generation of tile-level languages and compilers: Triton \cite{triton2019} from OpenAI, TileLang \cite{tilelang} from Peking University, adopted by DeepSeek as a primary development path, and Huawei's open-source PyPTO \cite{pypto}.

\textbf{PyPTO does not stop at the kernel.} It is also meant for programming the node, the SuperNode, and the cluster, and it does so in a modern programming style: Python syntax, with tiles and scopes as ordinary objects. A single program can therefore describe the behavior of a million processors, rather than the inner loop of one device.

What lets one program reach across those scales is what the runtime carries underneath: a nested machine model. The machine is presented as layers within layers, each with its own units, bandwidth, latency, and boundary. Because every layer is represented, every layer can be scheduled, and each layer can carry out the BSP of that layer: local work within the scope, exchange and aggregation among its units, and synchronization at its boundary. Since depth and radix change from one generation to the next, a runtime that holds the machine model this way can be retargeted rather than rewritten.

Three consequences follow, and each is what MPI + OpenMP cannot easily provide. One abstraction spans every level, because a tile is a smaller instance of the same kind of object and the nesting need not stop at the chip edge. Mapping becomes a decision rather than a rewrite: which layer a scope lands on is chosen against the machine model, so widening a tensor-parallel group from a board to a SuperNode is a remapping. And synchronization becomes a property of a scope rather than a different API per level, whether that scope sits inside a package or spans a data hall.

The correspondence with Nested BSP is close: a hierarchy of BSP computations under peer equality on one side, a hierarchy of schedulable scopes on the other. Practice will decide whether this is the right long-term expression, and other models may do it better. What survives either way is structural: when the programming model, the machine model, and the hardware nest all describe the same structure, the size of problem that can be attempted grows.

For AI, that already holds, since the workload tiles recursively at every level. For the recursively decomposable HPC workloads it should hold too: such a code would state its factorization, its V-cycle, or its elimination tree once, with scopes named, and let the placement of each phase follow the machine it runs on. Genuinely flat workloads fall back on one SuperNode, a ceiling that keeps rising because SuperNodes are growing to meet an AI industry far larger than HPC has ever been. Scientific computing may end up inheriting a tightly coupled machine bigger than it could ever have justified building on its own.

\section{Building It: Six Things the Unified Bus Had to Get Right}

With the principle established, the question returns to hardware: what actually builds this nested computer, where earlier technology fell short, and how the gap was closed.

The target comes first. Farther out means longer distance, lower bandwidth, and higher latency; that is physics, and it is accepted. The real failure of the past was not a gentle slope but a cliff. One step past the chassis, bandwidth dropped by an order of magnitude and latency rose by several. Inside the rack the system behaved like one machine; outside the rack, each boundary had its own protocol. The job of the critical technology is to pave that cliff into a slope, so that command passes down without a single layer dropped.

\textbf{1. One protocol, the same inside the rack and outside it.} The past used two. Inside the box, a bus was fast and nanosecond-class but born for short reach and unable to leave the chassis. Outside the box, a network could reach far, but every boundary was a toll booth: unpack, inspect, re-pack. One language inside the rack, another between racks; at every boundary, a transfer. In large clusters, more than 80\% of energy goes to moving data, and a large share of that is lost at these transfers \cite{unifiedbus}. The approach here merges ``bus'' and ``unified'' into one: a single protocol from the package all the way to the autonomous zone, end to end, with no transfers.

\textbf{2. Distant memory that feels like memory at hand.} Moving data used to mean climbing layer by layer to the application, then converting across protocols. A TCP/IP round trip cost tens of microseconds, mostly in software. RDMA improved that sharply, but the exchange was still, in essence, ``send a message, wait for a reply.'' Barriers and reductions were all constrained by the same pattern, and the larger the scale, the greater the penalty. The switch is to memory semantics: native load and store, with consistency handled in hardware. One communication round trip fell from tens of microseconds to around a hundred nanoseconds, a factor of roughly five hundred, and per-chip I/O bandwidth reaches the 8~Tbps class \cite{unifiedbus}.

\textbf{3. Master--slave torn down --- the foundation of peer equality.} The past was host commands, device obeys, everything through a center. That master--slave pattern is not an accident of one product line; it is the default grammar of traditional computer design: CPU as master, accelerators as slaves; host as master, devices as slaves; a control plane that owns initiation, and everyone else waiting. The larger the system, the more the center became a bottleneck; adding units did not add strength, and one plus one came to less than two. This design puts CPU, NPU, memory, storage, and NICs on the same bus as peers: anyone can initiate, anyone can respond. Barrier and aggregation at every Nested BSP layer need not report back to a center for every act. Without that equality, the nested plan collapses into a queue at the master; with it, the Nested Parallel von Neumann Architecture can scale without a single bottleneck through which everything must pass.

\textbf{4. Copper near, optics far --- with the language unchanged.} Copper has hit its wall. Push the rate of one wire up, and the wire grows thicker and reaches shorter; bundle thousands of copper cables and they become too thick to install. First-generation Unified Bus was therefore designed for both electrical and optical: copper nearby to protect latency, optics farther out to protect scale, with the protocol unchanged when the medium changes. On the optical side the choice was near-package optics, NPO. On a full electrical path the loss exceeds twenty decibels; NPO removes nine to eleven of them in one step, leaving three as all that co-packaged optics (CPO) can still take. What that buys is optical engines that can be built, tested, and replaced as single modules, latency at the ten-nanosecond class, and overall cost more than 40\% below CPO. This path is now an OIF project, walked by dozens of partners \cite{oif}.

\textbf{5. No cramming the soldiers into one tent.} The old instinct was to pack to the limit: larger chips, fuller racks. But compute grows with area, while I/O bandwidth and power grow only with perimeter, so the mismatch worsens as packing tightens. Heat is more concrete still: cooling a megawatt rack takes on the order of hundreds of square meters of heat-exchange area; a gigawatt hall may hold only a thousand racks yet occupy a square kilometer, with water pipes stretched a kilometer. Floor space in the hall is cheap; chips in the rack are expensive. Hence no chase after megawatt racks. The system stands open, bound by optics into one machine: stand sparsely, compute tightly; physically sparse, logically tight.

\textbf{6. In each generation, only a few hard battles.} Twenty physical limits taken on at once, each at ninety percent odds, leave a joint success rate of about one in eight. Five can still ship. Choosing what to take on and what to leave to a better form is itself part of design.

\section{In Practice: Three Things That Must Be Done Right}
\label{sec:practice}

Principle and path are set. Three things have to be right in what is built today.

\textbf{1. Switches must be high radix.} The more exits at a junction, the fewer handoffs across the system, so higher radix is the one item no one argues with. For the same scale, higher radix means fewer intermediate layers, and the latency and power saved are real.

\textbf{2. Compute chips must also be high radix.} This one is often overlooked. The old view held that the chip should just compute, with one or two outbound links considered enough and the rest left to the network. In a system of a million, that is like forcing a whole battalion's dispatches through a single door.

The chip itself must be a junction. More exits buy three layers of gain: higher total outbound bandwidth; more paths, so that if one breaks another remains (in ten-thousand-GPU clusters, link and module failure is routine); and, most important, one-hop coverage equal to switch ports times chip ports.

Both sides are multipliers. Double the chip's radix and the one-hop SuperNode scale doubles. The larger that one-hop reach, the shorter the path for barrier and aggregation, and it is exactly the innermost, most frequent Nested BSP layers that fall inside that reach. Raising the product of these two radices sets the parallelism at every nested hardware layer.

\textbf{3. High-density NPO --- light brought to the very edge of the compute chip.} The closer electro-optical conversion sits to the chip edge, the shorter the copper segment that remains. What this buys sounds contradictory but is natural: physically sparse, logically tight.

Logically tight: one-hop reach, memory semantics, the whole SuperNode still one computer. Physically sparse: chips need not crowd into a single chassis, and can spread out. Across this six-layer nest, physical space can open and relax freely; nothing needs to be crushed into a tiny volume.

\begin{figure}[htbp]
\centering
\includegraphics[width=0.7\linewidth]{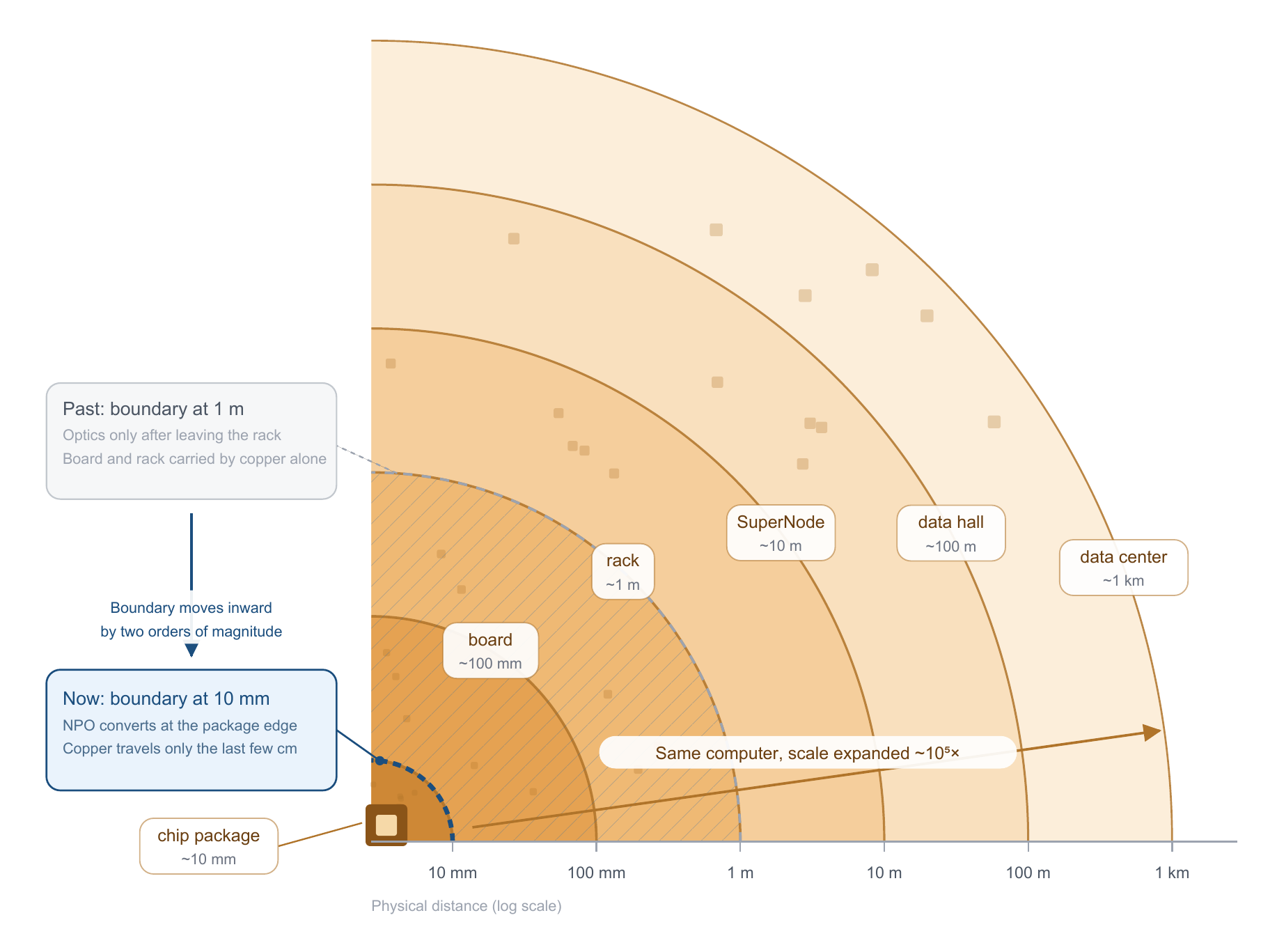}
\caption{Design freedom from NPO: package at ten millimeters, board at ten centimeters, rack at one meter, SuperNode at ten meters, data hall at a hundred meters, and data center at a kilometer. Moving the electro-optical boundary inward from 1~m to 10~mm relaxes every outer layer --- physically sparse, logically tight.}
\label{fig:scale}
\end{figure}

The scales span five orders of magnitude from a package to a data center (Fig.~\ref{fig:scale}).

The critical ring is the innermost. In the past, the electro-optical boundary sat around one meter: optics only after leaving the chassis. That meant the whole stretch from package through board to rack was carried by copper alone. Copper is unforgiving: push the rate up, and the wire grows thicker and reaches shorter; the board fights insertion loss, the rack fights density, and every step runs against a physical limit. NPO moves that boundary inward by two orders of magnitude, to ten millimeters at the package edge. Copper travels only the last few centimeters, and everything beyond is handed to light.

Once the boundary moves inward, every outer layer relaxes. The board no longer fights insertion loss for dozens of copper cables; the rack need not pack full or pile to a megawatt; racks can simply pull apart, because optical reach makes that distance immaterial. Each layer outward can expand tenfold in scale, with components sitting more sparsely. Through that entire expansion, the system remains logically one computer.

The limit of spatial density therefore need not be challenged at all. Cooling, power, and reliability no longer have to advance against physical limits at the same time. ``Stand sparsely, compute tightly'' comes down to this one move.

\section{Conclusion: Many Processors, Still One Computer}

The argument reduces to two extensions.

\textbf{Extend BSP to Nested BSP.} Turing formalized how a computation can be expressed as a program for a universal machine; von Neumann turned that programmability into the stored-program computer. BSP carried the programming question into the parallel domain by organizing local computation, communication, and synchronization into measurable supersteps. Nested BSP carries it further, into a hierarchical machine: an outer superstep contains smaller parallel computations, which may themselves contain still smaller ones. MPI, OpenMP, accelerator kernels, and collective libraries can implement these different scopes. Under peer equality, each layer coordinates locally before communicating outward; at every layer, the $\tau$ Scaling law folds time. One layer alone folds only so much, while six layers multiply the effect and bring down the time of the whole computation.

\textbf{Extend von Neumann to the Nested Parallel von Neumann Architecture, with the Unified Bus as the interconnect that lands it.} Von Neumann defined one computer; the Nested Parallel von Neumann Architecture extends that definition to how a formation of processors remains one computer: software and hardware nests aligned, memory semantics end to end, every node a peer on the bus, physically sparse and logically tight. It pairs with $\tau$: $\tau$ is the law of time folding, and the Nested Parallel von Neumann Architecture is the extended, peer-equal architecture that sets master--slave aside.

Put together, a SuperNode can stand more than eight thousand nodes, with aggregate memory bandwidth of 6.7~PB/s, full interconnect of 400~Tbps, and a full-formation barrier under ten microseconds. For users the effect is threefold: add chips and compute is added; chip strength is no longer spent on moving data; and the cost per token comes down.

Follow this nested structure outward and scale can keep growing. A single Super AI computer system at the 256K-node class is now being deployed. That is one machine, not a network of two hundred thousand machines glued together, but one computer with one Nested BSP plan and one Unified Bus command set through to the end.

The specification is open. A foundation of this kind must be larger than one company: whoever connects to the standard gets the same latency and the same linearity.

\textbf{Many processors, still one computer --- all peers.}

\end{document}